\documentclass[12pt]{iopart}

\usepackage{graphicx}
\usepackage{rotating}
\usepackage[pdftitle={Mass Dispersions in TDHF},pdfauthor={J. M. A. Broomfield and P. D. Stevenson},colorlinks=true,citecolor=black,linkcolor=black]{hyperref}

\begin{document}

\title[Mass dispersions from giant dipole resonances]
  {Mass dispersions from giant dipole resonances using the Balian-V\'{e}n\'{e}roni variational approach}

\author{J.~M.~A.~Broomfield and P.~D.~Stevenson}

\address{Dept.~of Physics, University of Surrey, Guildford, Surrey, GU2 7XH, United Kingdom}
\begin{abstract}
The Balian-V\'{e}n\'{e}roni variational approach has been implemented using a 3-dimensional time-dependent Hartree-Fock (TDHF)
code with realistic Skyrme interactions and used to investigate the mass dispersions from giant dipole resonances in $^{32}$S and
$^{132}$Sn decaying through particle emission. The fluctuations obtained are shown to be quantitatively larger than the standard
TDHF results.
\end{abstract}

\submitto{\JPG}
\maketitle
\section{Introduction}

The time-dependent Hartree-Fock (TDHF) approach can be used to
determine the expectation values of single-particle observables, such
as fragment mass, in nuclear reactions and decays but is known to
underestimate the fluctuations in these 
values \cite{cite:reinhard}. This is due to the 1-body nature of TDHF
and the fact that it neglects 2-body correlations. This problem has
previously been studied by Balian and V\'{e}n\'{e}roni
\cite{cite:bv1,cite:bv2,cite:bv3}, who derived a general variational
theory for the determination of expectation values, correlations and fluctuations. They found that, given the state of a system described, at the time $t_0$, by the 1-body density matrix,
$\rho\left(t_0\right)$ (a Slater determinant satisfying $\rho^2=\rho$),
the fluctuation, $\Delta Q$, in a 1-body observable, $Q$, at some later time $t_1$, is given by
\begin{equation}
\left. \left( \Delta Q_{BV} \right)^2 \right|_{t_1}
  = \lim\limits_{\varepsilon \to 0} \frac{1}{2\varepsilon^2}
     \mbox{Tr}\left[ \rho\left(t_0\right) - \sigma\left(t_0,\varepsilon\right) \right],
\label{eqn:bv}
\end{equation}
where $\sigma\left(t,\varepsilon\right)$ is a 1-body density matrix related to $\rho\left(t\right)$ through the boundary
condition
\begin{equation}
\sigma\left(t_1,\varepsilon\right)
  = \exp\left(\mbox{i}\varepsilon Q\right) \rho\left(t_1\right) \exp\left(-\mbox{i}\varepsilon Q\right), \label{eqn:sigma}
\end{equation}
and where the time evolution of $\rho\left(t\right)$ and $\sigma\left(t,\varepsilon\right)$ is given by the
usual TDHF equation. This result is significantly different from the standard TDHF result
\begin{eqnarray}
\left. \left( \Delta Q_{TDHF} \right)^2 \right|_{t_1}
  & = & \left. \left( \langle Q^2 \rangle - \langle Q \rangle^2 \right) \right|_{t_1}, \nonumber \\
  & = & \mbox{Tr} \left[ Q \rho\left(t_1\right) Q \left( 1 - \rho\left(t_1\right) \right) \right], \label{eqn:tdhf}
\end{eqnarray}
in that it depends on the initial time, $t_0$, with the final time, $t_1$, entering only through the boundary condition (\ref{eqn:sigma}). The
other key feature of this result is that it contains, through (\ref{eqn:sigma}), the observable $Q$ such that this
method is specifically tuned to the determination of the fluctuation of the observable of interest.

A practical implementation of (\ref{eqn:bv}) requires that a
Hartree-Fock calculation be performed to determine the initial state,
$\rho\left(t_0\right)$. The system is then excited by a suitable
external excitation, and a TDHF calculation performed from $t_0 \to
t_1$ to determine 
$\rho\left(t_1\right)$. This is used to obtain
$\sigma\left(t_1,\varepsilon\right)$ using (\ref{eqn:sigma}) and a
second TDHF calculation is then performed with the TDHF code run
backwards, $t_1 \to t_0$, to obtain
$\sigma\left(t_0,\varepsilon\right)$. The transformation
(\ref{eqn:sigma}) and the second TDHF calculation must
be repeated for a range of values of $\varepsilon$ to allow $\Delta
Q_{BV}$ to be determined by extrapolation to $\varepsilon\to 0$. 

The large number of computations required to evaluate (\ref{eqn:bv}) and the complexity of these calculations means that only a
handful of calculations have been performed using this method and those calculations which have been performed have used
simplified interactions and made use of symmetries (either spherical \cite{cite:troudet}, or axial
\cite{cite:marston,cite:bonche}) to render the problems tractable. However, modern advances in computing power mean that
this approach can now be implemented using fully 3-dimensional TDHF codes with full Skyrme interactions
\cite{cite:maruhn,cite:umar,cite:simenel,cite:nakatsukasa}.

We consider the mass dispersion in a bounded region of space around
a giant dipole resonance (GDR) which decays through particle emission and calculate the
mass (number of nucleons) in the nucleus according to
\begin{equation}
N\left(R_c\right)
  = \sum\limits_{m<\epsilon_F} \int \mbox{d}\bar{r} \left| \phi_m\left(\bar{r}\right) \right|^2 \theta\left(R_c - \left|\bar{r}\right|\right),
\label{eqn-n}
\end{equation}
where $R_c$ is the cutoff radius used to define the bounded region of space.

The nucleus was excited by multiplying the ground state wavefunctions from the HF calculation by a dipole boost given by
\begin{equation}
B_{D}\left(x,y,z\right)
  = \exp\left( \mbox{i} F C \left( A_x x + A_y y + A_z z\right) \right) \label{eqn:gdr-boost}
\end{equation}
with
\begin{equation}
C = \sqrt{\frac{5}{4\pi}} \frac{1}{1+\exp\left(\sqrt{x^2+y^2+z^2}\right)} \nonumber
\end{equation}
and where, for protons, $F = 1/Z$, and for neutrons, $F=-1/(A-Z)$,
where $A$ is the atomic mass number of the nucleus under investigation
and $Z$ is its charge. $A_x$, $A_y$ and $A_z$ determine the strength
of the boost applied to the nucleus.

Written in terms of the single particle wavefunctions (\ref{eqn:bv}) becomes \cite{cite:marston}
\begin{equation}
\left. \left( \Delta N_{BV} \right)^2 \right|_{t_1}
  = A - \lim\limits_{\varepsilon \to 0} \frac{f\left(\varepsilon\right)}{\varepsilon^2}, \label{eqn:sp-bv} \\
\end{equation}
\begin{equation*}
f\left(\varepsilon\right)
  = \sum\limits_{m,n<\epsilon_F} \int \mbox{d}\bar{r}
    \left| \left< \psi_m\left(t_0,\bar{r},\varepsilon\right) \right.\left| \phi_n\left(t_0,\bar{r}\right) \right> \right|^2.
\end{equation*}
The wavefunctions $\left|\phi_n\left(t\right)\right>$ were 
obtained from the results of a static Hartree-Fock calculation, 
whilst the wavefunctions
$\left|\psi_m\left(t,\bar{r},\varepsilon\right)\right>$ result
from the backwards TDHF calculations 
and are related to the wavefunctions
$\left|\phi_n\left(t,\bar{r}\right)\right>$ 
through the boundary condition
\begin{equation}
\psi\left(t_1,\bar{r},\varepsilon\right)
  = \exp\left(\mbox{i}\varepsilon \theta\left(R_c-\left|\bar{r}\right|\right)\right) \phi\left(t_1,\bar{r}\right).
\label{eqn:sp-psi}
\end{equation}

\section{GDR in \texorpdfstring{$^{32}$S}{32S}}

We consider first a GDR in $^{32}$S calculated using the Skyrme interaction with the SLy6 \cite{cite:chabanat}
parametrisation. All calculations were performed in a cubic model
space of size $32\times 32\times 32$ fm discretised in steps of  
$1$ fm. The initial HF calculation gave a $^{32}$S ground state with a total binding energy of $260.36$ MeV (compared
with the experimental value of $271.78$ MeV \cite{cite:audi}) and a prolate deformation ($\beta_2=0.11$).

At the beginning of the dynamic calculation the ground state wavefunctions were boosted in accordance with (\ref{eqn:gdr-boost}) and with
$A_x=A_y=A_z=112.5$ fm$^{-1}$. The simulation was allowed to run from an initial time $t_0=0$ fm/c to $t_1=250$ fm/c in steps of $0.2$ fm/c.
The emitted nucleons were reflected back from the boundary of the box and would, were the simulation allowed to run long enough,
re-enter the region occupied by the de-exciting nucleus causing
unphysical interactions. An analysis of the density and of $\langle N\left(R_c\right)\rangle$ as a function of time was
used to verify that the number of nuclei in the
nucleus had stabilised well in advance of the time $t_1$ and that the radiated flux had not
had enough time to be reflected back and to interact with the nucleus.

\begin{figure}
\begin{center}
\rotatebox{-90}{\includegraphics[width=.25\textwidth]{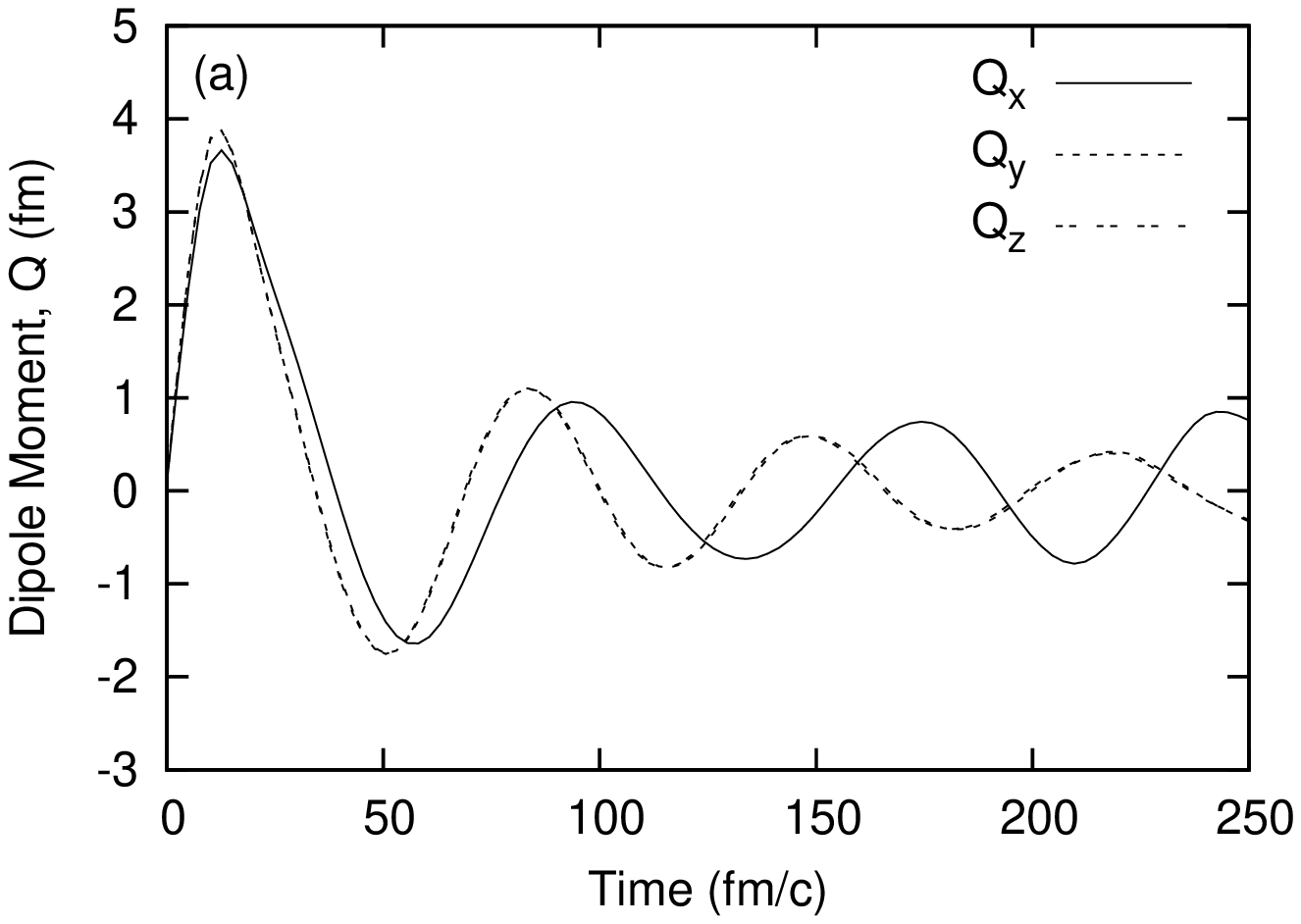}}
\rotatebox{-90}{\includegraphics[width=.25\textwidth]{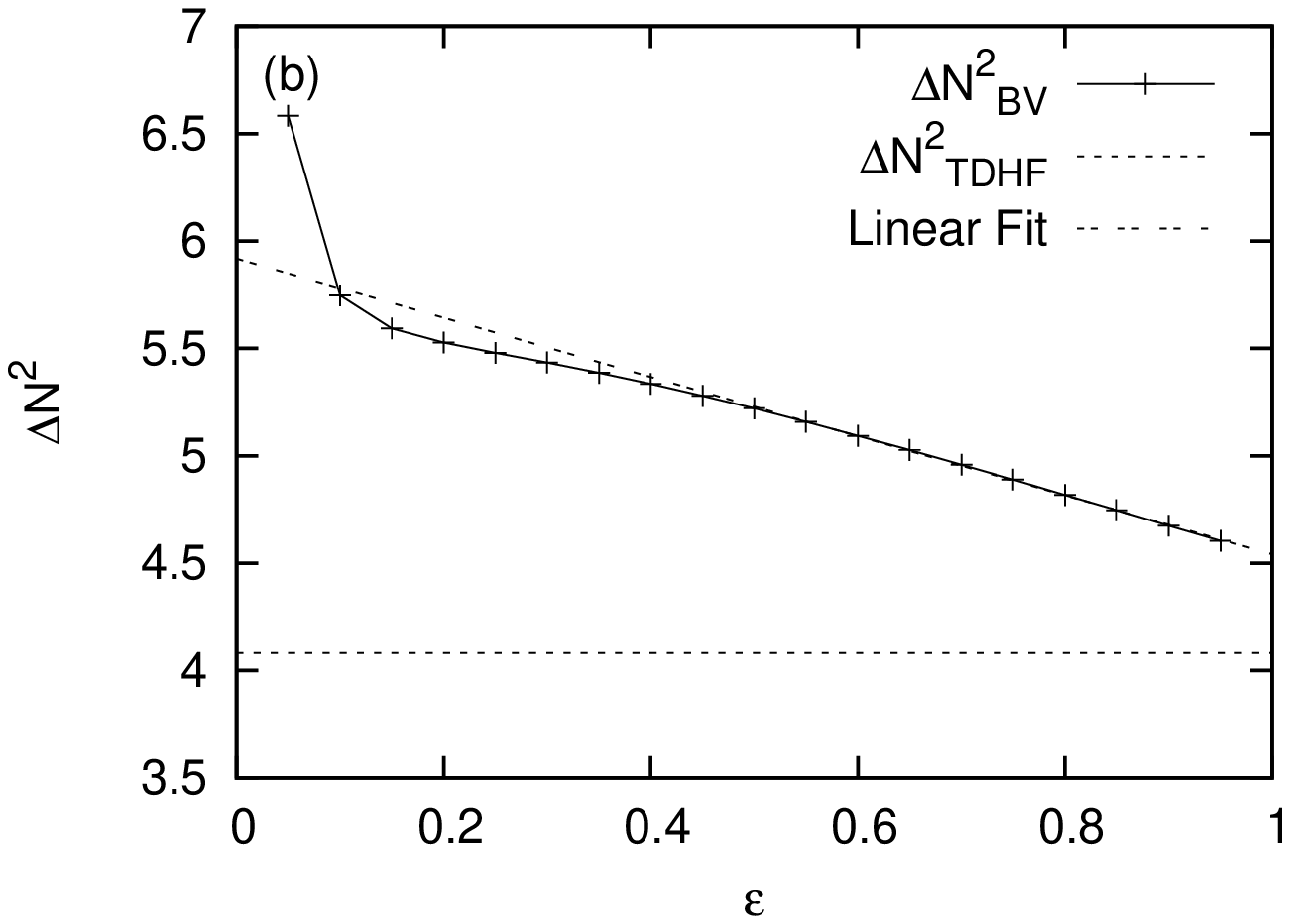}}
\end{center}
\caption
  {(a) The dipole moments ($Q_x$, $Q_y$ and $Q_z$) plotted as a function of time for a GDR in $^{32}$S. The difference between
  $Q_x$ and $Q_y$ and $Q_z$ is consistent with a calculation using a prolate deformed ground state where $x$ is the
  long axis. (b) $\Delta N^2_{BV}$ plotted as a function of $\varepsilon$ and extrapolated back to $\varepsilon=0$.
  The standard TDHF result (calculated at $t_1$ and independent of $\varepsilon$) is shown for reference.}
\label{fig:s32}
\end{figure}

The dipole moments, $Q_x$, $Q_y$ and $Q_z$, were obtained as a function of time using \cite{cite:simenel}
\begin{equation}
Q_i = \frac{\left(A-Z\right)Z}{A} \left( \langle x_i^P \rangle - \langle x_i^N \rangle \right),
\label{eqn:dip}
\end{equation}
where $i=1$,$2$,$3$ denotes $x$, $y$ and $z$ and $\langle x_i^P\rangle$ and $\langle x_i^N\rangle$ are the expectation values for
position calculated using the proton and neutron single particle states respectively. This is shown in figure \ref{fig:s32}(a). Due to the
prolate deformation of the $^{32}$S nucleus, 
the $Q_y$ and $Q_z$ values are identical and differ from the $Q_x$ values. The periodicity of $Q_x$, $Q_y$ and
$Q_z$ allow the excitation energies for the oscillations along each of
the three primary axes to be estimated. In this instance 
we obtain, for $Q_x$, a period of $\approx 71$ fm/c giving an excitation energy $E_x \approx 17.5$ MeV and, for $Q_y$ and $Q_z$, a
period of $\approx 68$ fm/c giving an excitation energy $E_y\approx E_z\approx 18.3$ MeV.

The final state gave $\langle N \rangle=26.65$ with $\Delta N_{TDHF}^2=4.08$ using $R_c=8$ fm which
represents the emission of $\approx 5$ nucleons. $R_c$ was chosen so that the bounded region fully enclosed the nucleus but
omitted, as much as possible, the extended (or dissipated) components of the wavefunctions. The transformation
(\ref{eqn:sp-psi}) was then applied and the TDHF code was run in reverse. This process was repeated for $\varepsilon$ values in
the range $0.05 \le \varepsilon \le 0.95$ in steps of $0.05$. At the end of each time-reversed calculation the fluctuation,
$\Delta N_{BV}^2\left(\varepsilon\right)$, was estimated using (\ref{eqn:sp-bv}). These values were plotted (see figure
\ref{fig:s32}(b)) and a straight line was fitted to the linear section of the graph and extended back to $\varepsilon=0$ to
obtain $\Delta N_{BV}^2=5.92$ which represents a $20$\% increase in $\Delta N$ using the BV approach compared with the standard
TDHF result. This graph is typical of those obtained using this approach and is linear for larger values of $\varepsilon$
increasing asymptotically as $\varepsilon\to 0$ due to the
$1/\varepsilon^2$ term in (\ref{eqn:bv}). Often, as in this case, the 
curve decreases for intermediate values of $\varepsilon$ where the reduced value of $\varepsilon$ means that the transformation
(\ref{eqn:sp-psi}) only has a small effect making the numerator in
(\ref{eqn:bv}) numerically approximately zero and dominant over the 
$\varepsilon^2$ denominator.

\section{GDR in \texorpdfstring{$^{132}$Sn}{132SN}}

\begin{figure}
\begin{center}
\rotatebox{-90}{\includegraphics[width=.25\textwidth]{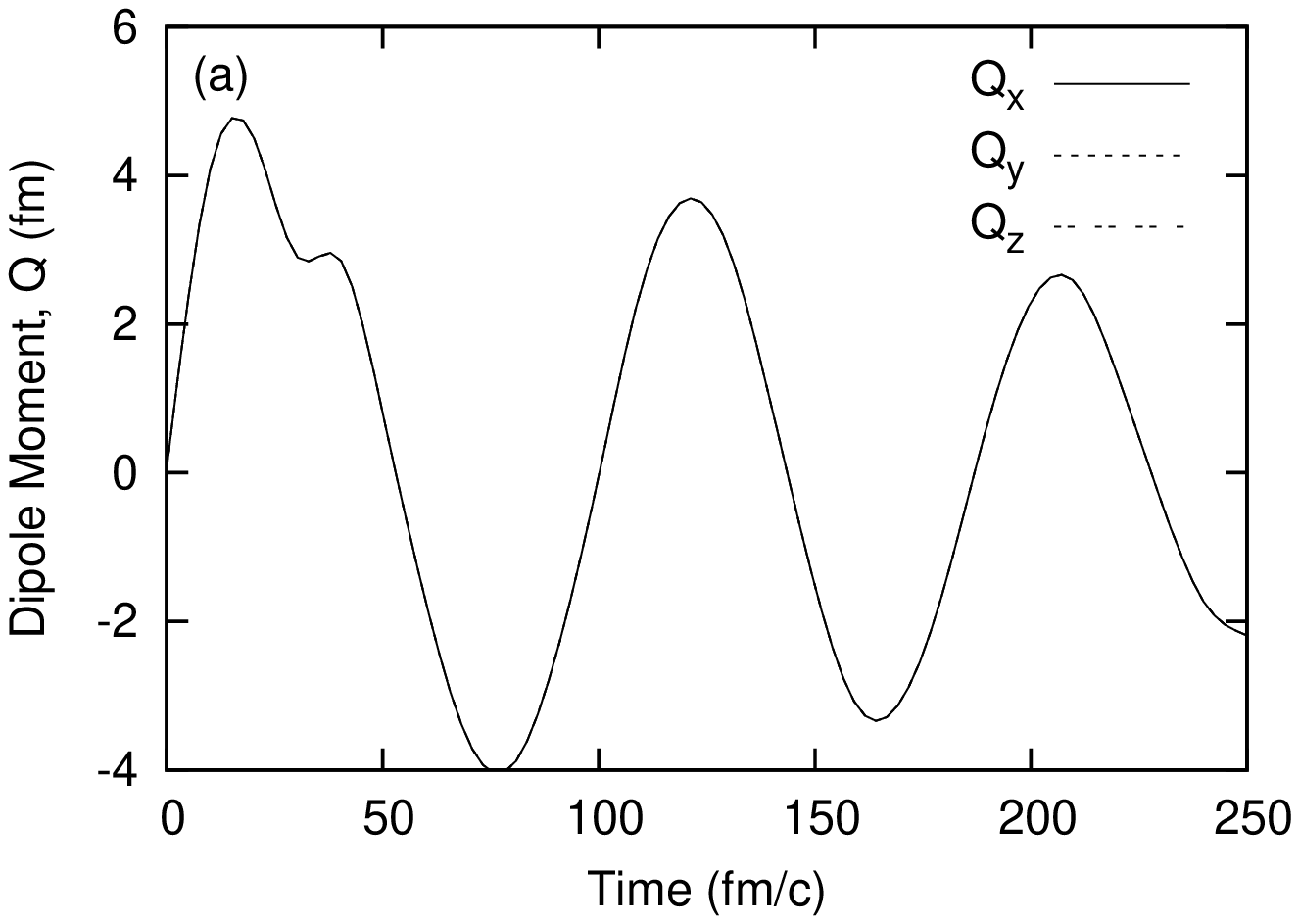}}
\rotatebox{-90}{\includegraphics[width=.25\textwidth]{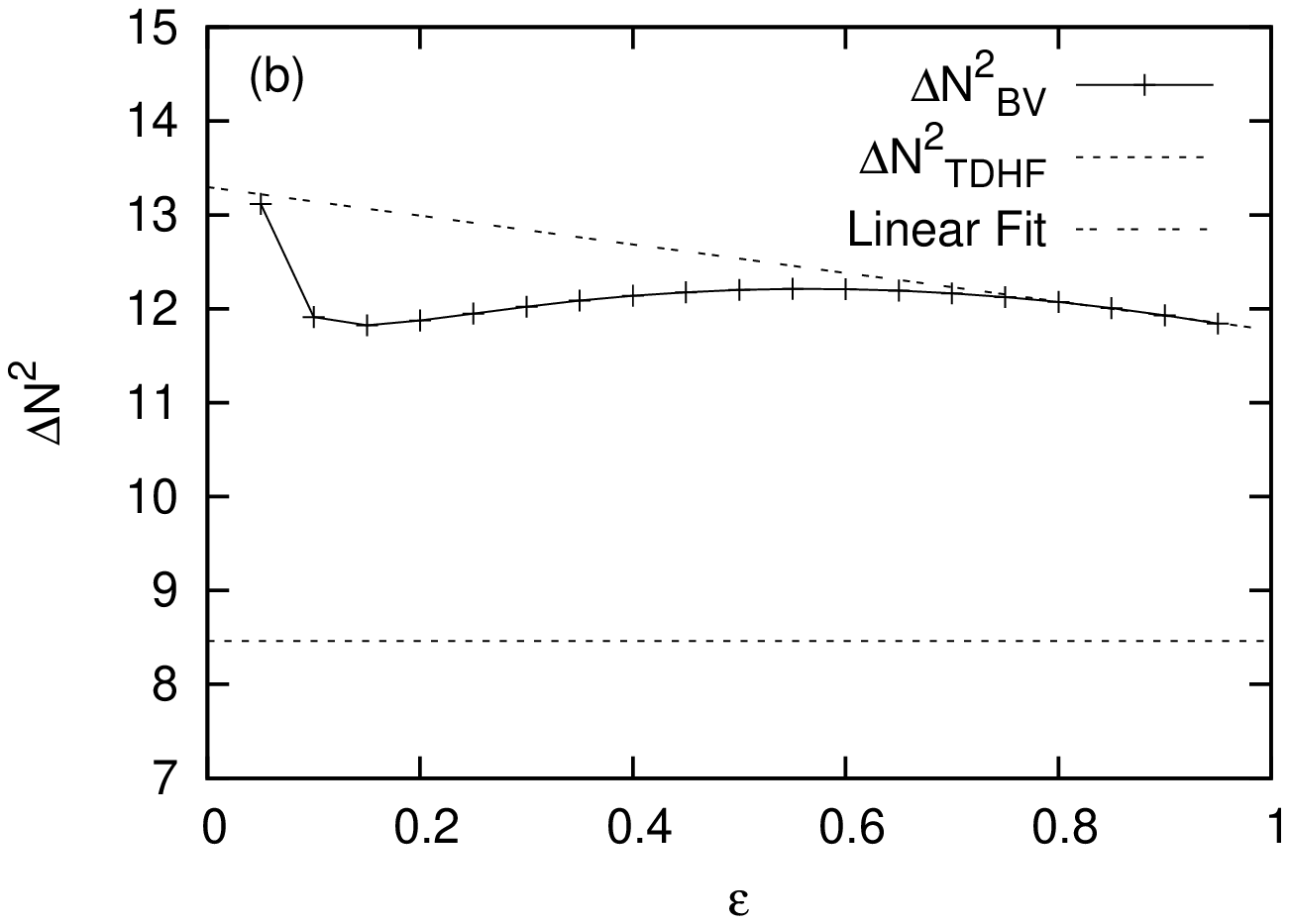}}
\end{center}
\caption
  {(a) The dipole moments ($Q_x$, $Q_y$ and $Q_z$) plotted as a function of time for a GDR in $^{132}$Sn.
  $Q_x$ and $Q_y$ and $Q_z$ as a result of the ground state being spherical.
  The shoulder at $\approx 40$ fm/c is a consequence of the $8$ fm cutoff radius.
  (b) $\Delta N^2_{BV}$ plotted as a function of $\varepsilon$ and extrapolated back to $\varepsilon=0$.
  The standard TDHF result (calculated at $t_1$ and independent of $\varepsilon$) is shown for reference.}
\label{fig:sn132}
\end{figure}

These calculations have been repeated for the doubly magic nucleus
$^{132}$Sn. All the calculations were carried out using the same
model space and interaction as the $^{32}$S calculation. The HF
calculation produced an undeformed ground state with a binding
energy of $1099.71$ MeV (compared with the accepted value of
$1102.85$ MeV \cite{cite:audi}). The ground state single particle
wavefunctions were boosted at the start of the TDHF calculation in
accordance with (\ref{eqn:gdr-boost}) and with $A_x=A_y=A_z=600$
fm$^{-1}$ and the calculation was run from $t_0=0$ fm/c to $t_1=250$
fm/c as in the previous calculation. The dipole moments were plotted
as a function of time and are shown in figure \ref{fig:sn132}(a).
The graph shows $Q_x$, $Q_y$ and $Q_z$ to be identical as expected
for a spherical nucleus and gives the periodicity of the dipole
moments as $\approx 88$ fm/c which corresponds to a resonance energy
of $\approx 14.1$ MeV. This is close to
the experimentally measured value of $16.1\left(7\right)$ MeV
\cite{cite:adrich}.

The standard THDF calculation gave, at the time $t_1$, $\langle N\rangle=121.02$ and $\Delta N_{TDHF}^{2}=8.46$ representing the
emission of 11 nucleons. A series of transformations and time-reversed TDHF calculations were carried out as previously. The
resulting graph, and linear fit, are shown in figure \ref{fig:sn132}(b) which gives $\Delta N_{BV}^{2}=13.30$ and represents a
$25$\% increase in $\Delta N$ compared with the standard TDHF result.

\section{Conclusions}

The Balian-V\'{e}n\'{e}roni approach has been implemented for the first time using a 3-dimensional TDHF code with the full Skyrme interaction.
Calculations have been performed for GDRs in $^{32}$S and $^{132}$Sn
and have demonstrated that the BV approach does produce
quantitatively larger results for the fluctuations of 1-body
operators. This approach is now being applied to heavy ion
collisions. 

\section*{Acknowledgements}

The authors are pleased to acknowledge the assistance of, and discussions with,
Ph.~Chomaz, R.~Balian, J.~S.~Al-Khalili and E.~B.~Suckling. This work was supported by EPSRC grant no.~EP/P501679/1.


\section*{References}

\begin{thebibliography}{11}
\bibitem{cite:reinhard}    
  P.-G.~Reinhard, R.~Y.~Cusson and K.~Goeke, Nuc.~Phys.~A{\bf 398}, 141-188 (1983)
\bibitem{cite:bv1}         
  R.~Balian and M.~V\'{e}n\'{e}roni, Annals of Physics {\bf 281}, 65-142 (2000)
\bibitem{cite:bv2}         
  R.~Balian and M.~V\'{e}n\'{e}roni, Annals of Physics {\bf 187}, 29-78 (1988)
\bibitem{cite:bv3}         
  R.~Balian and M.~V\'{e}n\'{e}roni, Annals of Physics {\bf 216}, 351-430 (1992)
\bibitem{cite:troudet}     
  T.~Troudet and D.~Vautherin, Phys.~Rev.~C {\bf 31}(1), 278-279 (1985)
\bibitem{cite:marston}     
  J.~B.~Marston and S.~E.~Koonin, Phys.~Rev.~Lett.~{\bf 54}(11), 1139-1141 (1985)
\bibitem{cite:bonche}      
  P.~Bonche and H.~Flocard, Nucl.~Phys.~A{\bf 437}, 189-207 (1985)
\bibitem{cite:maruhn}      
  J.~A.~Maruhn, P.-G.~Reinhard, P.~D.~Stevenson and M.~R.~Strayer, Phys.~Rev.~C {\bf 74}, 027601 (2006)
\bibitem{cite:umar}        
  A.~S.~Umar and V.~E.~Oberacker, Phys.~Rev.~C {\bf 74}, 024606 (2006)
\bibitem{cite:simenel}     
  C.~Simenel, Ph.~Chomaz and D.~de France, Phys.~Rev.~C {\bf 76}, 024609 (2007)
\bibitem{cite:nakatsukasa} 
  T.~Nakatsukasa and K.~Yabana, Nucl.~Phys.~A{\bf 788}, 349c-354c (2007)
\bibitem{cite:chabanat}    
  E.~Chabanat, P.~Bonche, P.~Haensel, J.~Meyer and R.~Schaeffer, Nucl.~Phys.~A{\bf 635}, 231-256 (1998)
\bibitem{cite:audi}        
  G.~Audi, A.~Wapstra and C.~Thibault, Nucl.~Phys.~A{\bf 729}, 337-676 (2003)
\bibitem{cite:adrich}      
  P.~Adrich {\it et al.}, Phys.~Rev.~Lett.~ {\bf 95}, 132501 (2005)
\end{thebibliography}
\end{document}